# The price dynamics of common trading strategies

J. Doyne Farmer[1] and Shareen Joshi[2]

A deterministic trading strategy can be regarded as a signal processing element that uses external information and past prices as inputs and incorporates them into future prices. This paper uses a market maker based method of price formation to study the price dynamics induced by several commonly used financial trading strategies, showing how they amplify noise, induce structure in prices, and cause phenomena such as excess and clustered volatility.

# 1. Introduction

## 1.1 Motivation

Under the efficient market hypothesis prices should instantly and correctly adjust to reflect new information. There is evidence, however, that this may not be the case: The

1. McKinsey Professor, Santa Fe Institute, 1399 Hyde Park Rd., Santa Fe 87501, jdf@santafe.edu.
2. Economics Department, Yale University, New Haven CT.

largest price movements often occur with little or no news (Cutler et al. 1989), price volatility is strongly temporally correlated (Engle 1982), short term price fluctuations are non-normal[3], and prices may not accurately reflect rational valuations (Campbell and Shiller 1988). This suggests that markets have nontrivial internal dynamics. Traders may be thought of as signal processing elements, that process external information and incorporate it into future prices. Insofar as individual traders use deterministic decision rules, they act as signal filters and transducers, converting random information shocks into temporal patterns in prices. Through their interaction they can amplify incoming noisy information, alter its distribution, and induce temporal correlations in volatility and volume.

This paper[4] investigates a simple behavioral model for the price dynamics of a few, common archetypal trading strategies. The goal is to understand the signal processing properties of these strategies, both separately and in concert. There are three groups of agents: *value investors* (or *fundamentalists)* who hold an asset when they think it is undervalued and short it when it is overvalued; *trend followers* (a particular kind of *technical trader,* or *chartist*), who hold an asset when the price has been going up and sell it when it has been going down; and *market makers*, who absorb fluctuations in excess demand, lowering the price when they have to buy and raising it when they have to sell. These are of course only a few of the strategies actually used in real markets. But they are known to be widely used (Keim and Madhaven 1995, Menkoff 1998), and understanding their influence on prices provides a starting point for more realistic behavioral models.

## 1.2 Relation to previous work

The first behavioral model that treats the dynamics of trend followers and value investors that we are aware of is due to Beja and Goldman (1980). Assuming linear trading rules for each type of trading, they showed that equilibrium is unstable when the fraction of trend followers is sufficiently high. A related model using nonlinear investment rules was introduced by Day and Huang (1990), who demonstrated that this could result in chaotic price series. The Beja and Goldman model was extended by Chiarella (1992), who made the trend following rule nonlinear. When the fraction of trend followers is sufficiently low, the equilibrium is stable, but when it exceeds a critical value it becomes unstable, and is replaced by a limit cycle. The excess demand of each trader type oscillates as the cycle is traversed, causing sustained deviations from the equilibrium price. This model was further enhanced by Sethi (1996), who studied inventory accumulation, cash flow, and the cost of information acquisition. He showed that for certain parameter settings the money of trend followers and value investors oscillates, and when trend followers dominate there are periods where the amplitude of price oscillations is large. Except for some remarks by Chiarella, this work is done in a purely deterministic setting.

Studies along somewhat different lines have been made by Lux et al. (1997, 1998, 1999), and also by Brock and Hommes (1997, 1998, 1999). Both study the effect of switching between trend following and value investing behavior. Brock and Hommes

3. See e.g. Mandelbrot (1963, 1997), Lux (1996), Mantegna and Stanley (1999).

4. Many of these results originally appeared in preprint form in Farmer (1998).

assume market clearing, and focus their work on the bifurcation structure and conditions under which the dynamics are chaotic. The Lux et al. papers use a disequilibrium method of price formation, and focus their work on demonstrating agreement with more realistic price series. They also assume a stochastic value process, and stress the role of the market as a signal processor. Bouchaud and Cont (1998) introduced a “Langevin model”, which is closely related to the work presented here[5]. These are not the only studies along these lines; for example, see Goldbaum (1999), or for brief reviews see LeBaron (1999b) or Farmer (1999).

The model discussed here was developed independently, and takes this study in a somewhat different direction. The cast of characters is expanded to include a market maker[6]. This makes it possible to study the dynamics of each trading strategy individually, i.e. in a market that includes only that strategy, the market maker, and noisy inputs. We characterize the noise amplification and price autocorrelations caused by each strategy. We investigate simple linear strategies analytically, and also present some numerical results for a heterogeneous market with more complicated nonlinear strategies.

This study also raises some issues about price formation. We assume the market maker is risk neutral, setting the price in response to received orders, without worrying about accumulated inventory. The market framework and price formation mechanism are similar to that of Kyle (1985). We show that this leads to problems with stability of equilibrium for simple value investing strategies. While these are ameliorated for some more complex value strategies, this study illustrates the importance of including market maker risk aversion in the price formation process.

## 2. Price formation model

In most real markets changes in the demand of individual agents are expressed in terms of *orders*. The two most common types of orders are market orders and limit orders. A *market order* is a request to transact immediately at the best available price. The fill price for small market orders is often quoted, so that it is known in advance, but for large market orders the fill price is unknown. In contrast, a *limit order* is a request to transact only at a given price or better. Thus the fill price is known, but the time of the transaction is unknown -- indeed the transaction may not be completed at all. In both cases there is uncertainty in either the time or the price of the transaction. Thus markets using either type of order violate the general equilibrium theory assumption of perfect competition.

5. Many of the results in this paper were presented at a seminar at Jussieu in Paris in June 1997. Bouchaud and Cont (1998) acknowledge their attendance in a footnote.

6. This was also done by Day and Huang (1990). Empirical studies of market clearing give half lives for market maker inventories on the order of a week (Hansch et al., 1998). This suggests that the lack of market clearing can be important on short timescales, e.g. a day.

## 2.1 Model framework

We will study only market orders. The goal of this section is to derive a *market impact function* $\phi$ *(*sometimes also called a *price impact function)* that relates the net of all such orders at any given time to prices. We assume there are two broad types of financial agents, trading a single asset (measured in units of shares) that can be converted to *money* (which can be viewed as a risk free asset paying no interest). The first type of agents are *directional traders.* They buy or sell by placing market orders, which are always filled. In the typical case that the buy and sell orders of the directional traders do not match, the excess is taken up by the second type of agent, who is a *market maker.* The orders are filled by the market maker at a price that is shifted from the previous price, by an amount that depends on the net order of the directional traders. Buying drives the price up, and selling drives it down. The market impact function $\phi$ is the algorithm that the market maker uses to set prices. This defines a price formation rule relating the net order to the new price.

Let there be $N$ directional traders, labeled by the superscript $i$, holding $x_t^{(i)}$ shares at time $t$. Although this is not necessary, for simplicity we assume synchronous trading at times $\ldots, t-1, t, t+1, \ldots$ . Let the position of the $i^{th}$ directional trader be a function $x_{t+1}^{(i)} = x^{(i)}(P_t, P_{t-1}, \ldots, I_t^{(i)})$, where $I_t^{(i)}$ represents any additional external information. The function $x^{(i)}$ can be thought of as the *strategy* or *decision rule* of agent $i$. The order $\omega_t^{(i)}$ is determined from the position through the relation

$$\omega_t^{(i)} = x_t^{(i)} - x_{t-1}^{(i)} . \qquad \text{(Eq 1)}$$

A single timestep in the trading process can be decomposed into two parts:

1. The directional traders observe the most recent prices and information at time $t$ and submit orders $\omega_{t+1}^{(i)}$.
2. The market maker fills all the orders at the new price $P_{t+1}$

To keep things simple, we will assume that the price $P_t$ is a positive real number, and that positions, orders, and strategies are anonymous. This motivates the assumption that the market maker bases price formation only on the *net order*

$$\omega = \sum_{i=1}^{N} \omega^{(i)} .$$

The algorithm the market maker uses to compute the fill price for the net order $\omega$ is an increasing function of $\omega$

$$P_{t+1} = f(P_t, \omega) . \qquad \text{(Eq 2)}$$

The fact that the new price depends only the current order, and not on the accumulated inventory of shares held by the market maker, implies that the market maker must be risk neutral.

## 2.2 Derivation of market impact function

An approximation of the market impact function can be derived by assuming that $f$ is of the form

$$f(P_t, \omega) = P_t \phi(\omega), \qquad \text{(Eq 3)}$$

where $\phi$ is an increasing function with $\phi(0) = 1$. Taking logarithms and expanding in a Taylor's series, providing the derivative $\phi'(0)$ exists, to leading order

$$\log P_{t+1} - \log P_t \approx \frac{\omega}{\lambda}. \qquad \text{(Eq 4)}$$

This functional form for $\phi$ will be called the *log-linear* market impact function. $\lambda$ is a scale factor that normalizes the order size, and will be called the *liquidity.* $\lambda$ is the order size that will cause the price to change by a factor of $e$, measured in units of shares.

For an equilibrium model the clearing price depends only on the current demand functions. In contrast, if prices are determined based on market impact using equation 3, in the general case they are path dependent. That is, the price at any given time depends on the starting price as well as the sequence of previous net orders. The log-linear rule is somewhere in between: The price change over any given period of time depends only on the net order imbalance during that time. In fact, we can show that this property implies the log-linear rule: Suppose we require that two orders placed in succession result in the same price as a single order equal to their sum, i.e.

$$f(f(P, \omega_1), \omega_2) = f(P, \omega_1 + \omega_2). \qquad \text{(Eq 5)}$$

By grouping orders pairwise, repeated application of equation 5 makes it clear that the price change in any time interval only depends on the sum of the net orders in this interval. Substituting equation 3 into equation 5 gives

$$\phi(\omega_1 + \omega_2) = \phi(\omega_1)\phi(\omega_2).$$

This functional equation for $\phi$ has the solution

$$\phi(\omega) = e^{\omega/\lambda}, \qquad \text{(Eq 6)}$$

which is equivalent to equation 4. Other possible solutions are $\phi(\omega) = 0$ and $\phi(\omega) = 1$, but neither of these satisfy the requirement that $\phi$ is increasing.

Note that this is similar to Kyle's (1985) model, with the important difference that in Kyle's model $P_{t+1} - P_t = \omega/\lambda$. The exponential form for $\phi$ used here guarantees that prices remain positive.

## 2.3 Dynamics

We can now write down a dynamical system describing the interaction between trading decisions and prices. Letting $p_t = \log P_t$, and adding a noise term $\xi_{t+1}$, equation 4 becomes

$$p_{t+1} = p_t + \frac{1}{\lambda}\sum_{i=1}^{N}\omega^{(i)}(p_t, p_{t-1}, \ldots, I_t) + \xi_{t+1}. \qquad \text{(Eq 7)}$$

To complete the model we need to make the functions $\omega^{(i)}$ used by agent $i$ explicit. Note that from equation 1, $\omega^{(i)}$ is automatically defined once the function $x^{(i)}$ is given

The addition of the random term $\xi_t$ can be interpreted in one of two ways: It can be thought of as corresponding to “noise traders”, or “liquidity traders”, who submit orders at random, in which case it should be divided by $\lambda$. Alternatively, it can be thought of as simply corresponding to random perturbations in the price, for example random information that affects the market maker’s price setting decisions. By using the form in equation 7 we take the latter interpretation.

The choice of a discrete time, synchronized trading process is a matter of convenience. We could alternatively have used an asynchronous process with random updating (which is also easy to simulate), or a continuous time Weiner process (which has advantages for obtaining analytic results). The discrete time mapping used here is convenient because it avoids conceptual problems associated with stochastic processes and makes causality very explicit. The time $\Delta t$ corresponding to a single iteration should be thought of as the timescale on which the fastest traders observe and react to the price, e.g. a minute to a day.

In this model the number of shares is conserved, i.e., every time an agent buys a share another agent loses that share. Thus the sum of all the agents’ positions are constant, providing the market maker’s position is included. All the agents, including the market maker, are free to take arbitrarily large positions, including net short (negative) positions. Thus, they are effectively given infinite credit. We have made considerable studies of wealth dynamics, but these are beyond the scope of this paper. Some preliminary results can be found in Farmer (1998).

## 2.4 Discussion of assumptions

Many assumptions have been made that deserve discussion. We think the most questionable is neglecting the market maker’s risk. However, we discuss the others first.

The log linear rule is simply the first term in a Taylor’s expansion, and is not intended as an accurate model of market impact. Indeed, several different empirical studies suggest that the shift in the logarithm of the price shift plotted against order size is a concave nonlinear function[7]. The log-linear rule is just a reasonable first order model, with the accuracy one would expect from the first term in a Taylor expansion. The derivation above implicitly assumes that market impact is permanent. That is, price changes caused by a net order at any given time persist until new net orders cause other changes. In contrast, if the market impact is temporary, price changes decay, even without new order flow. By letting the price be a real number we neglect the possibility of different prices for buying and selling, which occurs in real markets.

There is an implicit assumption that the market is symmetric in the sense that there is no *a priori* difference between buying and selling. Indeed, with any price formation rule that satisfies equation 5 buying and selling are inverse operations. (This is clear by letting $\omega_2 = -\omega_1$, which implies that $f^{-1}(P, \omega) = f(P, -\omega)$). This is reasonable for currency markets and many derivative markets, but probably not for most stock markets. The short selling rules in the American stock market are an example of a built-in asymmetry. From an empirical point of view for American equities the market impact of buying and selling are different, as observed by Chan and Lakonishok (1993, 1995). Such asymmetries can be taken into account by in terms of a different liquidity for buying and selling.

For the purposes of this study strategies will be fixed. This implies, for example, that profits are not reinvested. We study the price dynamics in the context where the noise inputs are stationary; this means that, providing the dynamics of equation 7 are stable, the returns are also stationary. Although this is unlikely to be the case for real price series, it is a useful simplifying assumption.

Last but not least: The assumption that the market impact function $\phi$ depends only on the net order $\omega$ does not take into account the market maker's risk aversion. Real market makers use their ability to manipulate the price to keep their inventory as small as possible. This makes the price formation process depend on the market maker's inventory[8].

The significance of this is made clearer by comparison to a standard disequilibrium price formation model originally proposed by Walras (Walker 1996). This model is used, for example, by Beja and Goldman (1980) and their descendents.

$$\frac{dp}{dt} = -\beta D(p, \ldots). \qquad \text{(Eq 8)}$$

$D$ is the aggregate excess demand, which depends on the price and possibly other variables, and $\beta$ is a constant that determines the rate of approach to equilibrium. The justification for this price formation rule is purely heuristic: It allows deviations from equilibrium, but providing $D = D(p)$ is a monotonic increasing function of price alone, results in a stable exponential approach to equilibrium.

Although this looks similar to the price formation model used here, it is actually quite different. To see this, assume the initial price clears the market, so that the market maker's starting position is zero. The market maker's position at subsequent times is then just $-D$,

7. For discussions of empirical evidence concerning market impact see Hausman and Lo (1992), Chan and Lakonishok (1993, 1995), Campbell et al. (1997), Torre (1997), and Keim and Madhaven (1999). Zhang (1999) has offered a heuristic derivation of a nonlinear market impact rule.

8. See, e.g. Huang and Stoll (1994).

where $D$ is the accumulated change in the net position of the directional traders. The net order $\omega$, however, is the net change in position. In a continuous time setting the model used here is thus like equation 8 with $D$ replaced by $dD/dt$ and $\beta$ replaced by $1/\lambda$. Like Kyle's (1985) model, price changes depend on received orders rather than the accumulated position of the market maker. To get an analogous model to equation 8 it would be necessary for price changes to depend *only* on the market maker's position at time $t+1$. In a future paper (involving other co-authors), we will report on a model of price formation exploring both these effects, which reduces to either as a special case. But for now we will analyze the effects of the model developed here, which because of its similarity to Kyle's influential market making model has some credibility in the literature.

# 3. Agent behaviors

We now describe some trading strategies in more detail and study their price dynamics. Since the market maker is in a sense a "neutral" agent, we can begin by studying each strategy trading against the market maker. Each strategy induces price dynamics that characterize its signal processing properties. We can then study the noise amplification, autocorrelation function, and frequency response of each strategy.

One approach to classifying financial trading strategies is based on their information inputs. Decision rules that depend only on the price history are called *technical* or *chartist* strategies. *Trend following* strategies are a commonly used special case in which positions are positively correlated with recent price changes. *Value* or *fundamental* strategies, in contrast, are based on external information leading to a subjective assessment of the long term fundamental value. Investors using these strategies do not believe this is the same as the current price. Pure technical strategies can be thought of as signal filters: They accept past prices as inputs and transform future prices. Value strategies, in contrast, are primarily signal transducers: they use external value signals as inputs and, through their trading, incorporate them into prices.

## 3.1 Trend followers

*Trend followers,* also sometimes called positive-feedback investors (DeLong et al. 1990), invest based on the belief that price changes have inertia. A trend strategy takes a positive (long) position if prices have recently been going up, and a negative (short) position if they have recently been going down. More precisely, a trading strategy is trend following on timescale $\theta$ if the position $x_t$ has a positive correlation $\rho$ with past price movements on timescale $\theta$, i.e.

$$\rho(x_{t+1}, (p_t - p_{t-\theta})) > 0 .$$

A strategy can be trend following on some timescales but not on others.

An example of a simple linear trend following strategy, which can be regarded as a first order Taylor approximation of a general trend following strategy, is

$$x_{t+1} = c(p_t - p_{t-\theta}), \qquad \text{(Eq 9)}$$

where $c > 0$. Note that if we let $c < 0$ this becomes a *contrarian strategy.* Letting the l*og-return* $r_t = p_t - p_{t-1}$, from equation 7, the induced dynamics are

$$r_{t+1} = \alpha(r_t - r_{t-\theta}) + \xi_t. \qquad \text{(Eq 10)}$$

where $\alpha = c/\lambda > 1$ and $r_{t-\theta} = p_t - p_{t-\theta}$. Figure 1 shows a series of prices with $\alpha = 0.2$ and $\theta = 10$.

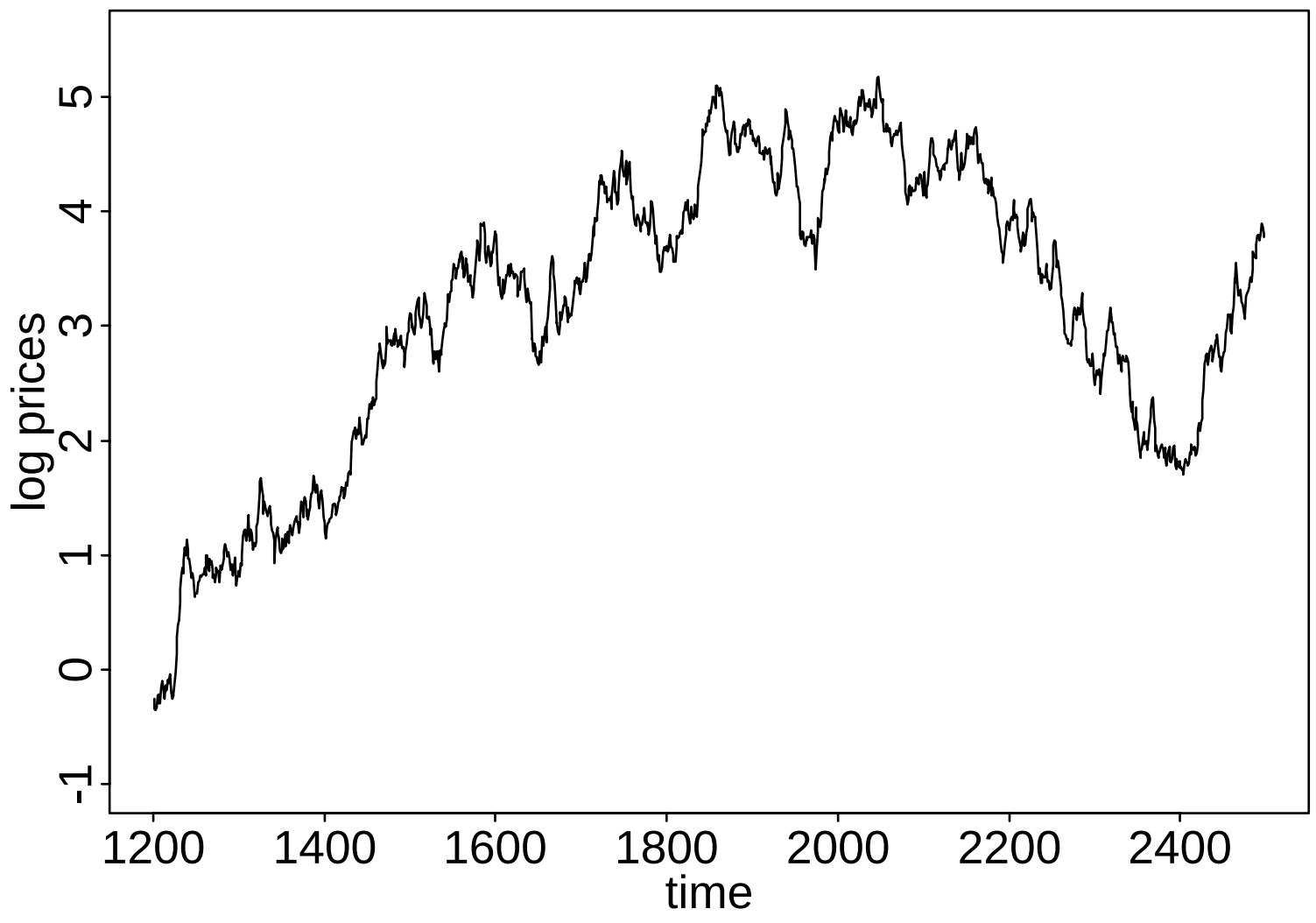

**FIGURE 1. Log price vs. time for trend followers with $\alpha = 0.2$ and $\theta = 10$ in Equation 10. Trend followers tend to induce short term trends in prices, but they also cause oscillations on longer timescales.**

The stability of the dynamics can be calculated by writing equation 10 in the form $u_{t+1} = Au_t$, where $u_t = (r_t, \ldots, r_{t-\theta})$, and computing the eigenvalues of $A$. For $\theta = 1$ these are

$$\varepsilon_{\pm} = \frac{\alpha(1-\alpha) \pm \sqrt{5 - 2\alpha + \alpha^2}}{2}.$$

The dynamics are stable when $\alpha < 1$.

Trend strategies overall amplify the noise in prices. This is reflected in the variance of the log-returns, which is computed by taking the variance of both sides of equation 10.

$$\sigma_r^2 = \frac{\sigma_\xi^2}{(1 - 2\alpha^2(1 - \rho_r(\theta)))}.$$

$\sigma_r^2$ is the variance of log-returns, and $\sigma_\xi^2$ is the variance of the noise $\xi_t$. Since the auto-correlation of log-returns, $\rho_r(\theta) \le 1$, it follows that $\sigma_r > \sigma_\xi$. Regardless of the value of $\alpha$ or $\rho_r$, the variance of the price fluctuations is larger than of the noise driving term. However, note that this is also true for a contrarian strategy: Reversing the sign of $c$ in equation 9 leaves this result unchanged. Thus we see that either trend or contrarian strategies can contribute to excess volatility by amplifying noise in prices.

Trend strategies induce trends in the price, but as we show below, they can also have other side effects. For example, consider Figure 2, which shows the autocorrelation function for the return series of Figure 1. The decaying oscillations between positive and negative values are characteristic of trend strategies with large lags. For $\tau = 1$ the autocorrelation function is of order $\alpha$. As $\tau$ increases it decays, crossing zero at roughly $\tau \approx \theta/2 + 1$. As $\tau$ continues to increase it becomes negative, reaching a minimum at $\tau = \theta + 1$, where it is of order $-\alpha$. The autocorrelation then increases again, reaching a local maximum at $\tau = 2\theta + 2$, where it is of order $\alpha^2$. As $\tau$ increases still further it oscillates between positive and negative values with period $2\theta + 2$, decaying by a factor of $\alpha$ with every successive period.

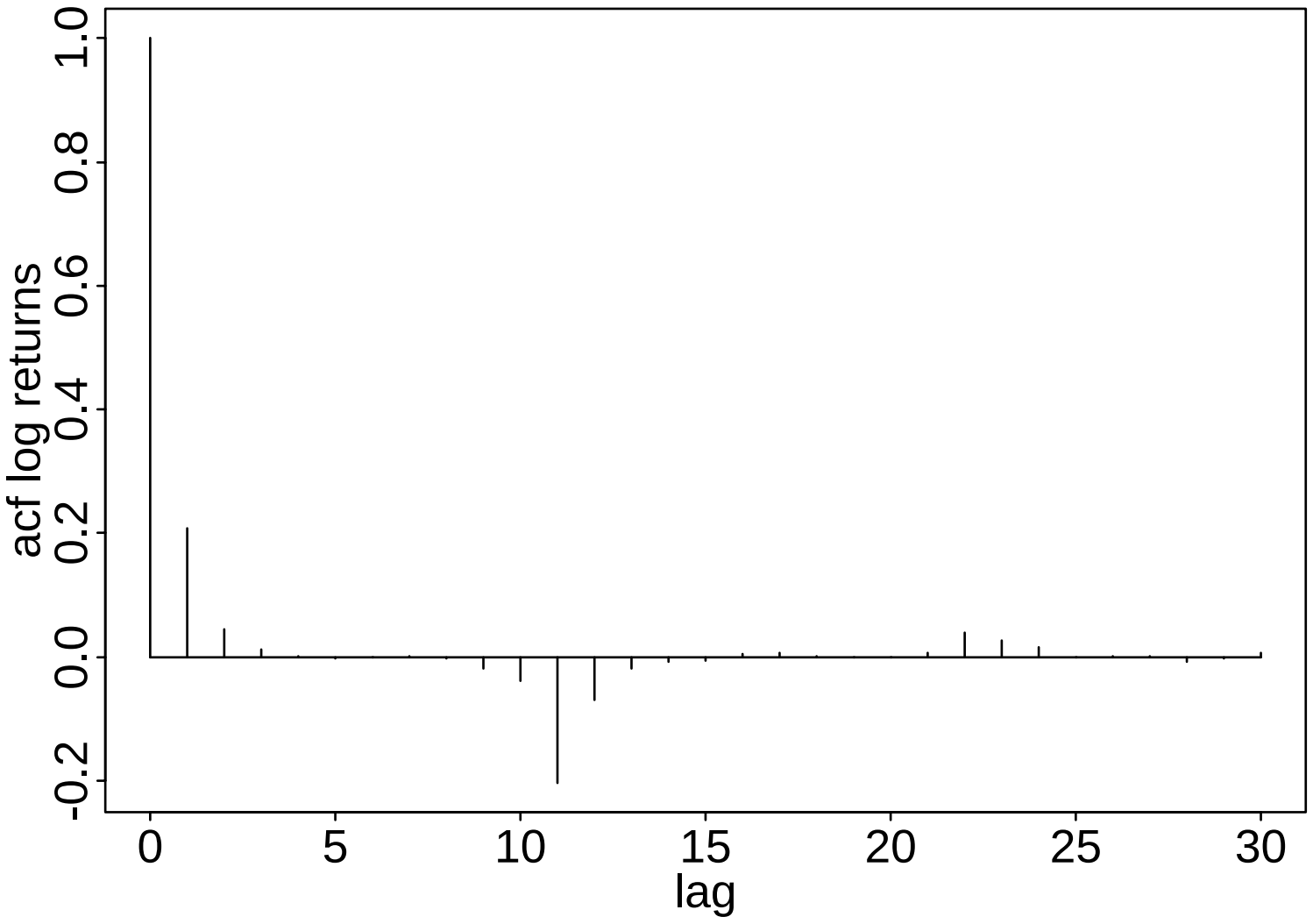

**FIGURE 2. The autocorrelation function for Equation 10 with $\alpha = 0.2$ and $\theta = 10$. The positive coefficients for small $\tau$ indicate short term trends in prices, and the negative coefficients indicate longer term oscillations.**

This behavior can be understood analytically. A recursion relation for the autocorrelation function can be obtained by multiplying equation 10 by $r_{t-n}$, subtracting the mean, and averaging, which gives

$$\rho_r(n+1) = \alpha(\rho_r(n) - \rho_r(|n-\theta|)) . \quad \text{(Eq 11)}$$

Doing this for $n = 0, \ldots, \theta - 1$ gives a system of $\theta$ linear equations that can be solved for the first $\theta$ values of $\rho_r(\tau)$ by making use of the requirement that $\rho_r(0) = 1$. The remainder of the terms can be found by iteration. For example, for $\theta = 1$, for $\tau = 1, \ldots, 6$ the autocorrelation function is

$$\rho_r(\tau) = \frac{1}{1+\alpha}(\alpha, -\alpha, -2\alpha^2, \alpha^2(1-2\alpha), \alpha^3(3-2\alpha), \alpha^4(3-2\alpha)). \quad \text{(Eq 12)}$$

Solving this for a few other values of $\theta$ demonstrates that the first autocorrelation $\rho_r(1)$ is always positive and of order $\alpha$, but $\rho_r(\theta+1)$ is always negative of order $-\alpha$. For large $\theta$ and small $\alpha$, using equation 11 it is easy to demonstrate that the autocorrelation follows the behavior described above. For $\tau \le \theta + 1$, to leading order in $\alpha$, $\rho(\tau) \approx \alpha^{\tau} - \alpha^{|\tau - \theta - 1| + 1}$.

Representing this in frequency space adds insight into the signal processing properties. The power spectrum of the returns can be computed by taking the cosine transform of the autocorrelation function, or alternatively, by computing the square of the Fourier transform of the log-returns and averaging. The result is shown in Figure 3. We see that the power spectrum has a large peak at frequency $2\theta + 2$, with peaks of decreasing amplitude at the odd harmonics of this frequency. The amplitude of the peaks is greater than one, indicating that the trend strategy amplifies these frequencies. However, the troughs, which occur at the even harmonics, have amplitude less than one, indicating that the trend strategy damps these frequencies. When viewed as a signal processing element, the trend strategy is essentially a selective low frequency noise amplifier, which induces oscillations at frequencies related to the time horizon over which trends are evaluated. The detailed properties are specific to this particular trend rule; in particular, the oscillations in the spectrum are caused by the fact that this trend rule uses a moving average with a sharp cut off. However, the basic property of amplifying low frequency noise is present in all the trend rules we have studied.

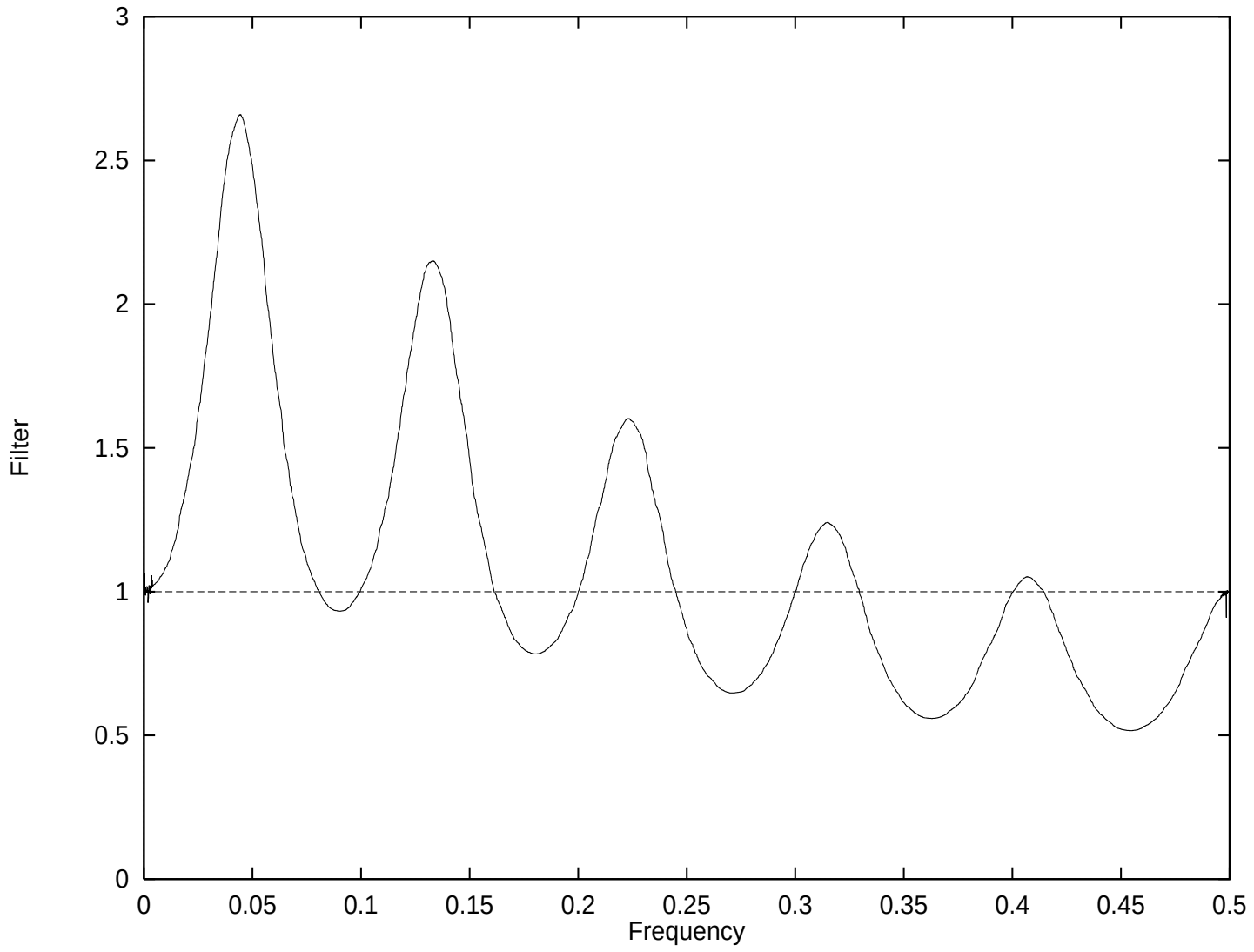

**FIGURE 3. Power spectrum of returns induced by a trend following strategy. Same parameters as Figure 2.**

## 3.2 Value investors

Value investors make a subjective assessment of value in relation to price. They believe that their perceived value may not be fully reflected in the current price, and that future prices will move toward their perceived value. They attempt to make profits by taking positive (long) positions when they think the market is undervalued and negative (short) positions when they think the market is overvalued.

In a homogeneous equilibrium setting in which everyone agrees on value, price and value are the same. In a nonequilibrium context, however, prices do not instantly reflect values - there can be interesting dynamics relating the two. Indeed many authors, such as Campbell and Shiller (1988), have suggested that prices may not track rational valuations very well, even in liquid markets, and that in some cases the differences can be substantial.

For the purposes of this paper it doesn't matter how individual agents form their opinions about value[9]. We take the estimated value as an exogenous input, and focus on the response of prices to changes in it. Let the logarithm of the value $v_t$ be a random walk,

$$v_{t+1} = v_t + \eta_{t+1}, \qquad \text{(Eq 13)}$$

where $\eta_t$ is a normal, IID noise process with standard deviation $\sigma_\eta$ and mean $\mu_\eta$. We will begin by studying the case where everyone perceives the same value, and return to study the case where there are diverse views about value in Section 3.2.5.

The natural way to quantify whether price tracks value is by using the concept of cointegration, introduced by Engle and Granger (1987). This concept is motivated by the possibility that two random processes can each be random walks, even though on average they tend to move together and stay near each other. More specifically, two random processes $y_t$ and $z_t$ are *cointegrated* if there is a linear combination $u_t = ay_t + bz_t$ that is stationary. For example, the log price and log value are cointegrated if $p_t - v_t$ has a well defined mean and standard deviation.

### 3.2.1 Simple value strategies

For the simplest class of value strategies the position is of the form

$$x_{t+1} = x(v_t, p_t) = V(v_t - p_t), \qquad \text{(Eq 14)}$$

9. They could, for example, use a standard dividend discount model, in which case their valuations depend on their forecasts of future dividends and interest rates. The results here, however, are independent of the method of valuation.

where $V$ is an increasing function with $V(0) = 0$, $v_t$ is the logarithm of the perceived value, and $p_t$ is the logarithm of the price. This class of strategies only depends on the *mispricing* $m_t = p_t - v_t$. Such a strategy takes a positive (long) position when the asset is underpriced; if the asset becomes even more underpriced, the position either stays the same or gets larger. Similarly, if the mispricing is positive it takes a negative (short) position.

If $V$ is differentiable we can expand it in a Taylor series. To first order the position can be approximated as

$$x_{t+1} = c(v_t - p_t),$$

where $c > 0$ is a constant proportional to the trading capital. From equations (1) and (7) the induced price dynamics in a market consisting only of this strategy and the market maker are

$$\begin{aligned} r_{t+1} &= -\alpha r_t + \alpha\eta_t + \xi_{t+1} \\ p_{t+1} &= p_t + r_{t+1} \end{aligned} . \qquad \text{(Eq 15)}$$

where $r_t = p_t - p_{t-1}$, $\eta_t = v_t - v_{t-1}$, and $\alpha = c/\lambda$. These dynamics are second order. This is evident from equation 15 since $p_{t+1}$ depends on both $p_t$ and $p_{t-1}$. The stability can be determined by neglecting the noise terms and writing equation 15 in the form $u_{t+1} = Au_t$, where $u_t = (r_t, p_t)$. The eigenvalues of $A$ are $(1, -\alpha)$. Thus when $\alpha \le 1$ the dynamics are neutrally stable, which implies that the logarithm of the price, like the logarithm of the value, follows a random walk. When $\alpha > 1$ the dynamics are unstable.

Simple value strategies induce negative first autocorrelations in the log-returns $r_t$. This is easily seen by multiplying both sides of equation 15 by $r_{t-i}$, subtracting the mean, and taking time averages. Assuming stationarity, this gives the recursion relation $\rho_r(\tau) = -\alpha\rho_r(\tau - 1)$. Since $\rho_r(0) = 1$, this implies

$$\rho_r(\tau) = (-\alpha)^{\tau}, \qquad \text{(Eq 16)}$$

where $\tau = 0, 1, 2, \ldots$. Because $\alpha > 0$, the first autocorrelation is always negative. Since the autocorrelation is determined by the linear part of $V$, this is true for any differentiable value strategy in the form of equation 14.

This value strategy amplifies the price noise $\xi_t$, but may or may not amplify the value noise $\eta_t$. To see this, compute the variance of the log-returns by squaring equation 15 and taking time averages. This gives

$$\sigma_r^2 = \frac{\alpha^2\sigma_\eta^2 + \sigma_\xi^2}{1 - \alpha^2}, \qquad \text{(Eq 17)}$$

where $\sigma_\eta^2$ and $\sigma_\xi^2$ are the variances of $\eta_t$ and $\xi_t$. This amplifies the external noise, since for any value of $\alpha$, $\sigma_r > \sigma_\xi$. Similarly, if $\alpha > 1/\sqrt{2}$ then $\sigma_r > \sigma_\eta$.

This strategy by itself does not cause prices to track values. This is evident because Equation 15 shows no explicit dependence on price or value. The lack of cointegration can be shown explicitly by substituting $m_t = p_t - v_t$ into Equation 15, which gives

$$\Delta m_{t+1} = -\alpha \Delta m_t - \eta_t + \xi_t,$$

where $\Delta m_t = m_t - m_{t-1}$. When $\alpha < 1$, $\Delta m_t$ is stationary and $m_t$ is a random walk. We have made several numerical simulations using various nonlinear forms for $V$, and we observe similar results. The intuitive reason for this behavior is that, while a trade entering a position moves the price toward value, an exiting trade of the same size moves it away from value by the same amount. Thus, while the negative autocorrelation induced by simple value strategies might reduce the rate at which prices drift from value, this is not sufficient for cointegration. The lack of cointegration can lead to problems with unbounded positions, implying unbounded risk. This comes about because the mispricing is unbounded, and the position is proportional to the mispricing. Thus if this is the only strategy present in the market the position is also unbounded. This problem disappears if another strategy is present in the market that cointegrates prices and values.

So far we have assumed ongoing changes in value. It is perhaps even more surprising that the price fails to converge even if the value changes once and then remains constant. To see this, consider equation 15 with $\Delta v_1 = v$, and $\Delta v_t = 0$ for $t > 1$. Assume $\xi_t = 0$, and for convenience let $p_1 = v_1 = 0$ and $\Delta p_1 = 0$. Iterating a few steps by hand shows that $p_t = (\alpha - \alpha^2 + \alpha^3 + \ldots(-\alpha)^{t-1})v$. If $\alpha < 1$, in the limit $t \to \infty$ this converges to $p_\infty = \alpha v/(1+\alpha)$. Thus when $\alpha < 1$ the price initially moves toward the new value, but it never reaches it; when $\alpha > 1$ the dynamics are unstable.

### 3.2.2 When do prices track values?

How can we solve the problem of making prices track values? One approach is to change the price formation rule. As already discussed in Section 2.4, this can be achieved by including risk aversion for the market maker. An alternative that is explored here is to investigate alternative value investing strategies. The order based value strategies discussed below fix the problem, but at the unacceptable cost of generating unbounded inventories. The threshold value strategies introduced in the following section manage to achieve both.

### 3.2.3 Order-based value strategies

One way to make prices track values is to make the strategy depend on the order instead of the position. A strategy of this type buys as long as the asset is underpriced, and sells as long as it is overpriced. Under the simple value strategy of the preceding section, if the mispricing reaches a given level, the trader takes a position. If the mispricing holds that level, he keeps the same position. For an order-based strategy, in contrast, if the asset is underpriced he will buy, and if on the next time step it is still mispriced he will buy again, and continue doing so as long as the asset remains mispriced. One can define an *order based value strategy* of the form

$$\omega_{t+1} = \omega(v_t, p_t) = W(v_t - p_t)$$

where as before $W$ is an increasing function with $W(0) = 0$. If we again expand in a Taylor's series, then to leading order this becomes

$$\omega_{t+1} = c(v_t - p_t).$$

Without presenting the details, let us simply state that it is possible to analyze the dynamics of this strategy and show that the mispricing has a well-defined standard deviation. Prices track values. The problem is that the position is not stationary, and the trader can accumulate an unbounded inventory. This is not surprising, given that this strategy does not depend on position.

The signal is to buy or sell as long as a mispricing persists, which means that typically the position is not forced to go to zero, even when the mispricing goes to zero. This problem occurs even in the presence of other strategies that cause cointegration of price and value. Numerical experiments suggest that non-linear extensions have similar problems. Real traders have risk constraints, which mean that position is of paramount concern. Strategies that do not depend on the position are unrealistic.

### 3.2.4 State-dependent threshold value strategies

The analysis above poses the question of whether there exist strategies that cointegrate prices and values and have bounded risk at the same time. This section introduces a class of strategies with this property.

From the point of view of a practitioner, a concern with the simple position-based value strategies of Section 3.2.1 is excessive transaction costs. Trades are made whenever the mispricing changes. A common approach to ameliorate this problem and reduce trading frequency is to use state dependent strategies, with a threshold for entering a position, and another threshold for exiting it. Like the simpler value strategies studied earlier, such strategies are based on the belief that the price will revert to the value. By only entering a position when the mispricing is large, and only exiting when it is small, the goal is to trade only when the expected price movement is large enough to beat transaction costs.

An example of such a strategy, which is both nonlinear and state dependent, can be constructed as follows: Assume that a short position $-c$ is entered when the mispricing exceeds a threshold $T$ and exited when it goes below a threshold $\tau$. Similarly, a long position $c$ is entered when the mispricing drops below a threshold $-T$ and exited when it exceeds $-\tau$. This is illustrated in Figure 4. Since this strategy depends on its own position as well as the mispricing, it can be thought of as a finite state machine, as shown in Figure 5.

In general different traders will choose different entry and exit thresholds. Let trader $i$ have entry threshold $T^{(i)}$ and exit threshold $\tau^{(i)}$. For the simulations presented here we will assume a uniform distribution of entry thresholds ranging from $T_{min}$ to $T_{max}$, and a uniform density of exit thresholds ranging from $\tau_{min}$ to $\tau_{max}$, with a random pairing of entry and exit thresholds. $c$ is chosen so that $c = a(T - \tau)$, where $a$ is a positive constant[10].

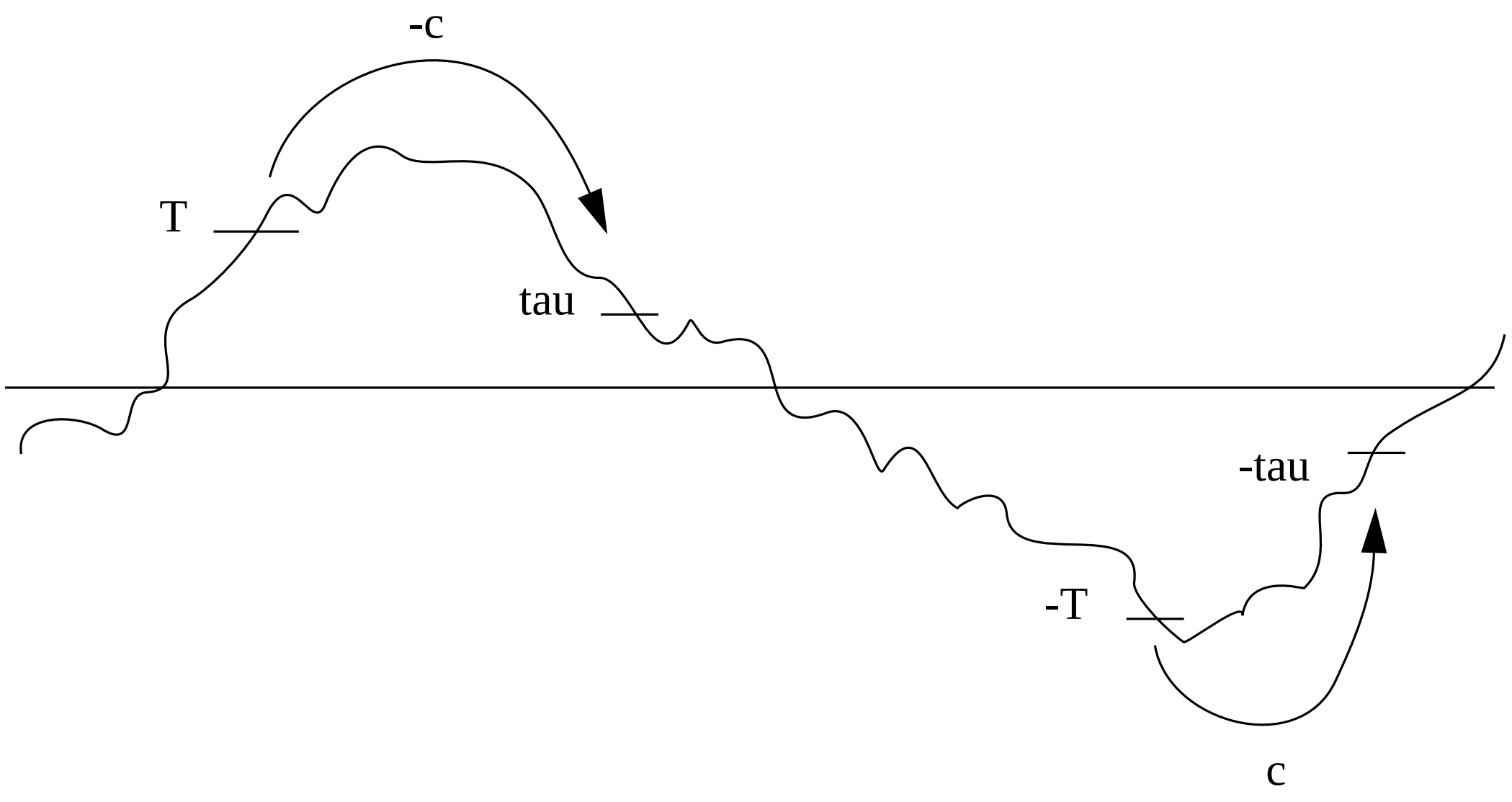

**FIGURE 4. Schematic view of a nonlinear, state-dependent value strategy. The trader enters a short position $-c$ when the mispricing $m_t = p_t - v_t$ exceeds a threshold $T$, and holds it until the mispricing goes below $\tau$. The reverse is true for long positions.**

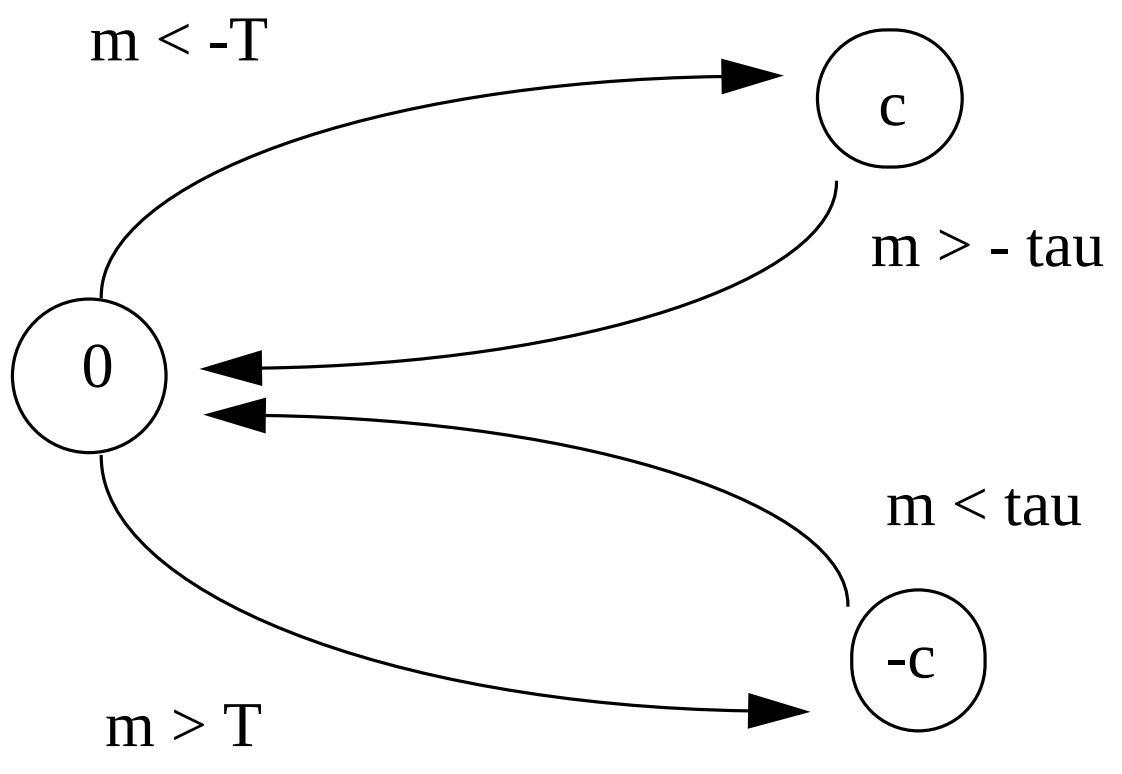

**FIGURE 5. The nonlinear state-dependent value strategy represented as a finite-state machine. From a zero position a long-position $c$ is entered when the mispricing $m$ drops below the threshold $-T$. This position is exited when the mispricing exceeds a threshold $-\tau$. Similarly, a short position $-c$ is entered when the mispricing exceeds a threshold $T$ and exited when it drops below a threshold $\tau$**

There are several requirements that must be met for this to be a sensible value strategy. The entry threshold should be positive and greater than the exit threshold, i.e. $T > 0$ and $T > \tau$. In contrast, there are plausible reasons to make $\tau$ either positive or negative. A trader who is very conservative about transaction costs, and wants to be sure that the full return has been extracted before the position is exited, will take $\tau < 0$. However, others might decide to exit their positions earlier, because they believe that once the price is near the value there is little expected return remaining. We can simulate a mixture of the two approaches by making $\tau_{min} < 0$ and $\tau_{max} > 0$. However, to be a sensible value strategy, a trader would not exit a position at a mispricing that is further from zero than the entry point. $\tau_{min}$ should not be *too* negative, so we should have $-T < \tau < T$ and $|\tau_{min}| \leq T_{min}$.

$\tau < 0$ is a desirable property for cointegration. When this is true the price changes induced by trading always have the opposite sign of the mispricing. This is true both entering and exiting the position. A simulation with $\tau_{max} = 0$ and $\tau_{min} < 0$ is shown in Figure 6. Numerical tests clearly show that the price and value are cointegrated. The cointegration is weak, however, in the sense that the mispricing can be large and keep the same sign for many iterations.

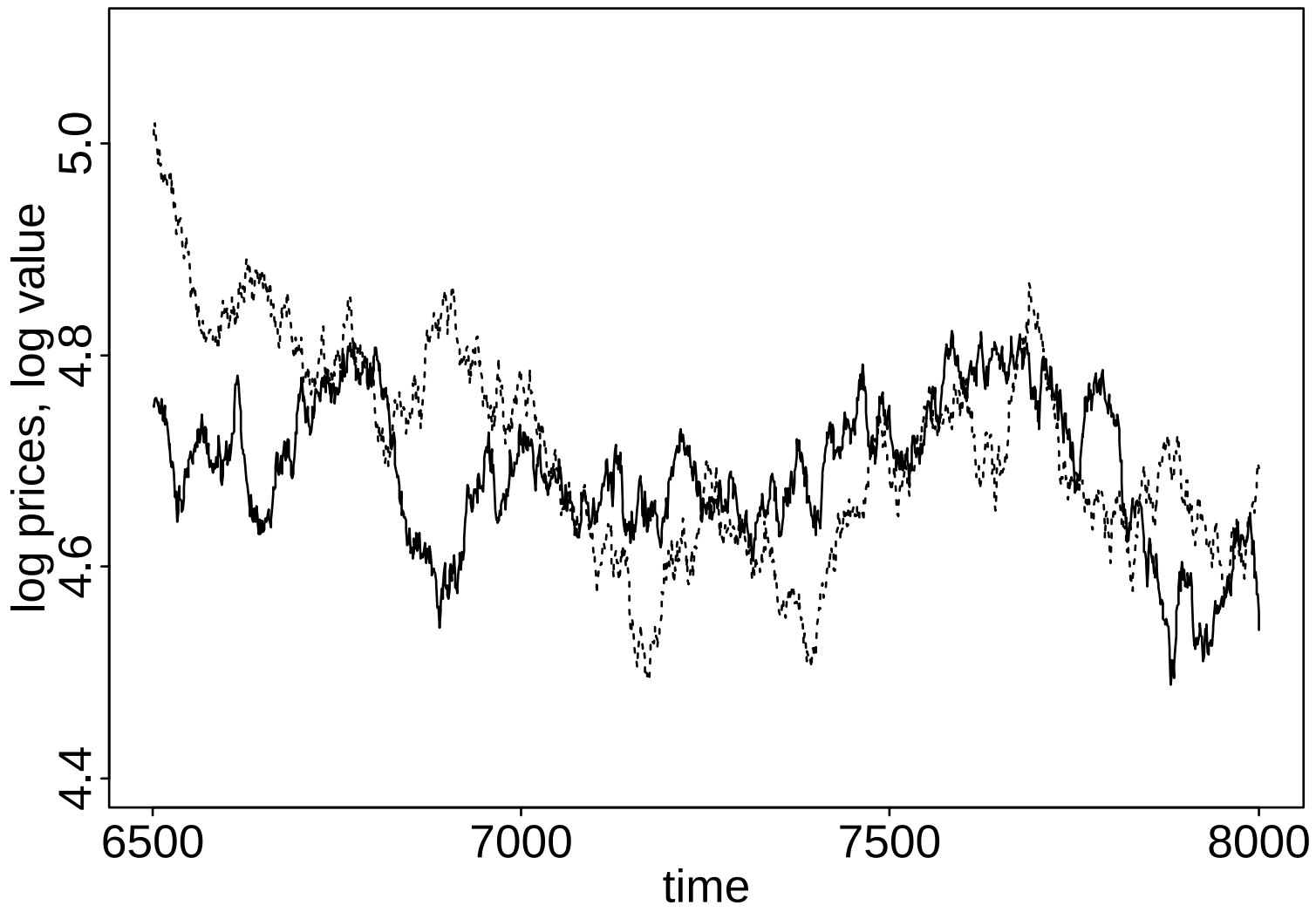

**FIGURE 6. The induced price dynamics of a nonlinear state-dependent value strategy with 1000 traders using different thresholds. The log-price is shown as a solid line and the log-value as a dashed line.** $\tau_{min} = -0.5$, $\tau_{max} = 0$, $T_{min} = 0.5$, $T_{max} = 6$, $N = 1000$, $a = 0.001$, $\sigma_\eta = 0.01$, **and** $\sigma_\xi = 0.01$, **and** $\lambda = 1$.

Figure 7 shows a simulation with the range of exit thresholds chosen so that $\tau_{min} < 0$ but $\tau_{max} > 0$. For comparison with Figure 6 all other parameters are the same. The price and value are still cointegrated, but more weakly than before. This is apparent from the increased amplitude of the mispricing. In addition, there is a tendency for the price to "bounce" as it approaches the value. This is caused by the fact that when the mispricing

10. This assignment is natural because traders managing more money (with larger $c$) incur larger transaction costs. Traders with larger positions need larger mispricings to make a profit.

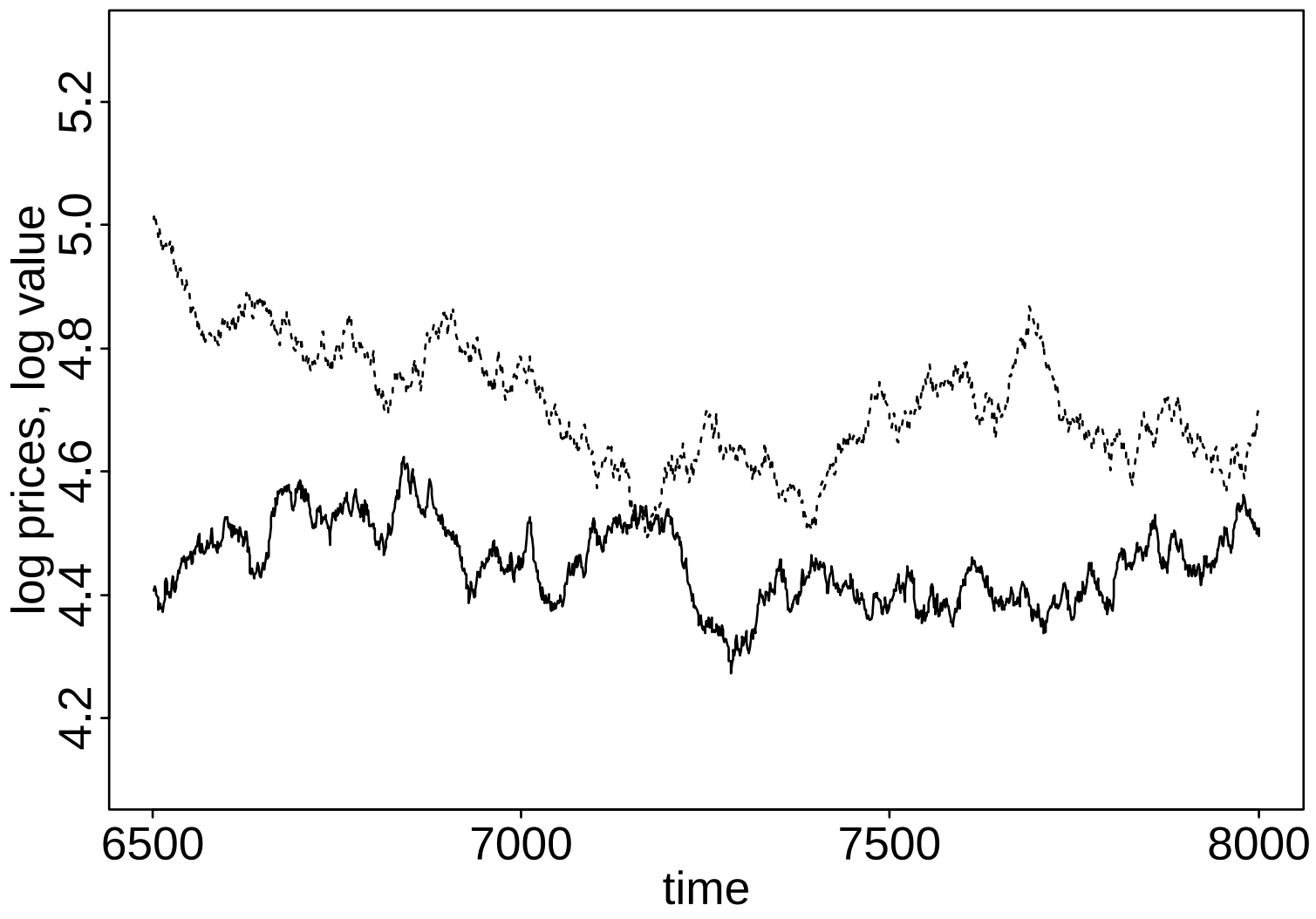

**FIGURE 7. Price (solid) and value (dashed) vs. time for the nonlinear state-dependent strategy of Figure 5. The parameters and random number seed are the same as Figure 6, except that $\tau_{min} = -0.5$ and $\tau_{max} = 0.5$**

approaches zero some traders exit their positions, which pushes the price away from the value. The value becomes a "resistance level" for the price (see e.g. Edwards and Magee, 1992), and there is a tendency for the mispricing to cross zero less frequently than it does when $\tau^{(i)} < 0$ for all $i$. Thus, we see that a value strategy can create patterns that could be exploited by a technical strategy. Based on results from numerical experiments it appears that the price and value are cointegrated as long as $\tau_{min} < 0$. Necessary and sufficient conditions for cointegration deserve further study[11].

### 3.2.5 Heterogeneous values, representative agents, and excess volatility

So far we have assumed a single perceived value, but given the tendency of people to disagree, in a more realistic setting there will be a spectrum of different values. We will show that in this case, for strategies that are linear in the logarithm of value, the price dynamics can be understood in terms of a single *representative agent*, whose perceived value is the mean of the group. However, for nonlinear strategies this is not true -- there exists no representative agent, and diverse perceptions of value can cause excess volatility.

Suppose there are $N$ different traders perceiving value $v_t^{(i)}$, using a value strategy $V^{(i)}(v_t, p_t) = c^{(i)}V(v_t, p_t)$, where $c^{(i)}$ is the capital of each individual strategy. The dynamics are

11. Problems can occur in the simulations if the capital $c = a(T - \tau)$ for each strategy is not assigned reasonably. If $a$ is too small the traders may not provide enough restoring force for the mispricing; once all $N$ traders are committed to a long or short position, price and value cease to be cointegrated. If $a$ is too big instabilities can result because the price kick provided by a single trader creates oscillations between entry and exit. Nonetheless, between these extremes there is a large parameter range with reasonable behavior.

$$p_{t+1} = p_t + \frac{1}{\lambda}\sum_{i=1}^{N} c^{(i)} V(v_t^{(i)}, p_t),$$

Providing the strategy is linear in the value the dynamics will be equivalent to those of a single agent with the average perceived value and the combined capital. This is true if $V$ satisfies the property

$$\sum_{i=1}^{N} c^{(i)} V(v_t^{(i)}, p_t) = cV(\bar{v}_t, p_t),$$

where

$$\bar{v}_t = \frac{1}{c}\sum_{i=1}^{N} c_i v_t^{(i)},$$

and $c = \sum c_i$. For example, the linearized value strategy of Section 3.2.1 satisfies this property. Thus, for strategies that depend linearly on the logarithm of value, the mean is sufficient to completely determine the price dynamics, and the diversity of opinions is unimportant. The market dynamics are those of a single representative agent.

The situation is quite different when the strategies depend nonlinearly on the value. To demonstrate how this leads to excess volatility, we will study the special case where traders perceive different values, but these values change in tandem. This way we are not introducing any additional noise to the value process by making it diverse, and any amplification in volatility clearly comes from the dynamics rather than something that has been added. The dynamics of the values can be modeled as a simple reference value process $\bar{v}_t$ that follows equation 13, with a fixed random offset $v^{(i)}$ for each trader. The value perceived by the $i^{th}$ trader at time $t$ is

$$v_t^{(i)} = \bar{v}_t + v^{(i)}. \qquad \text{(Eq 18)}$$

In the following simulations the value offsets are assigned uniformly between $v_{min}$ and $v_{max}$, where $v_{min} = -v_{max}$, so that range is $2v_{max}$.

We will define the excess volatility as

$$\varepsilon = \sqrt{\sigma_r^2/(\sigma_\eta^2 + \sigma_\xi^2)}, \qquad \text{(Eq 19)}$$

i.e. as the ratio of the volatility of the log-returns to the volatility of the exogenous noise. This measures the noise amplification. If $\varepsilon > 1$ the log-returns of prices are more volatile than the fluctuations driving the price dynamics. Figure 8 illustrates how the excess volatility increases as the diversity of perceived values increases, using the threshold value strategy of Section 3.2.4. The excess volatility also increases as the capital increases. This is caused by additional trading due to disagreements about value. In the linear case these would cancel and leave no effect on the price, but because of the nonlinearity of the strat-egy, this is not the case. If the market is a machine whose purpose is to keep the price near the value, this machine is noisy and inefficient.

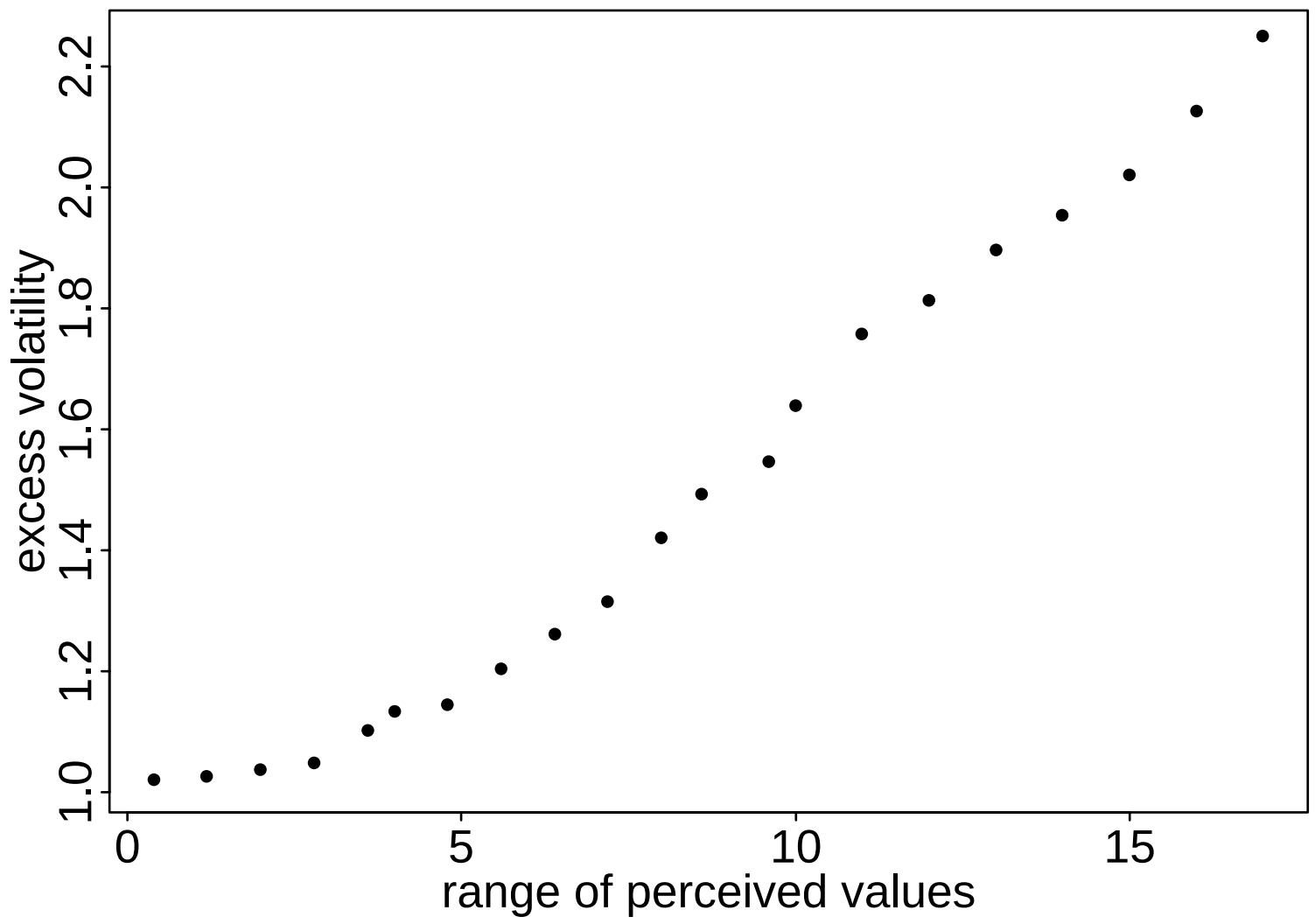

**FIGURE 8. Excess volatility as the range of perceived values increases while the capital is fixed at $0.035$. See Equation 19. The other parameters are the same as those in Figure 6.**

### 3.3 Value investors and trend followers together

In this section we investigate the dynamics in a more heterogeneous setting including both nonlinear trend following and value investing strategies. We use the threshold value strategies described in Section 3.2.4, and use the trend strategy of Section 3.1, except that we make it nonlinear by adding entry and exit thresholds, just as for the value strategy of Section 3.2.4. We make a qualitative comparison to annual prices and dividends for the S&P index[12] from 1889 to 1984, using the average dividend as a crude measure of value, and simulating the price dynamics on a daily timescale. As a proxy for daily value data we linearly interpolate the annual logarithm of the dividends, creating 250 surrogate trading days for each year of data. These provide the reference value process $\bar{v}_t$ in equation 18.

12. See Campbell and Shiller (1988). Both series are adjusted for inflation.

The parameters for the simulation are given in Table 1. There were two main criteria for choosing parameters: First, we wanted to match the empirical fact that the correlation of the log-returns is close to zero. This was done by matching the population of trend followers and value investors, so that the positive short term autocorrelation induced by the trend followers is cancelled by the negative short term autocorrelation of the value investors. Thus the common parameters for trend followers and value investors are the same. Second, we wanted to match the volatility of prices with the real data. This is done primarily by the choice of $a$ and $N$ in relation to $\lambda$, and secondarily by the choice of $v_{min}$ and $v_{max}$. Finally, we chose what we thought was a plausible timescale for trend following uniformly distributed from $1 – 100$ days.

**TABLE 1. Parameters for the simulation with trend followers and value investors in Figure 10.**

| Description of parameter | symbol | value |
|---|---|---|
| number of agents | $N_{value}$, $N_{trend}$ | 1200 |
| minimum threshold for entering positions | $T^{value}_{min}$, $T^{trend}_{min}$ | 0.2 |
| maximum threshold for entering positions | $T^{value}_{max}$, $T^{trend}_{max}$ | 4 |
| minimum threshold for exiting positions | $\tau^{value}_{min}$, $\tau^{trend}_{min}$ | –0.2 |
| maximum threshold for exiting positions | $\tau^{value}_{max}$, $\tau^{trend}_{max}$ | 0 |
| scale parameter for capital assignment | $a_{value}$, $a_{trend}$ | $2.5 \times 10^{-3}$ |
| minimum offset for log of perceived value | $v_{min}$ | –2 |
| maximum offset for log of perceived value | $v_{max}$ | 2 |
| minimum time delay for trend followers | $\theta_{min}$ | 1 |
| maximum time delay for trend followers | $\theta_{max}$ | 100 |
| noise driving price formation process | $\sigma_{\xi}$ | 0.35 |
| liquidity | $\lambda$ | 1 |

The real series of American prices and values are shown in Figure 9 and the simulation results are shown in Figure 10. There is a qualitative correspondence. In both series the price fluctuates around value, and mispricings persist for periods that are sometimes measured in decades. However, at this point no attempt has been made to make forecasts, which is not trivial for this kind of model. The point of the above simulation is just to demonstrate how a combination of trend and value investors results in oscillations in the mispricing.

Because of the choice of parameters there is no short term linear autocorrelation structure in this price series. There is plenty of nonlinear structure, however, as illustrated in Figure 11, which shows the smoothed volume[13] of value investors and trend followers as a function of time. The two groups of traders become active at different times, simply because the conditions that activate their trading are intermittent and unsynchronized. This is true even though the capital of both groups is fixed. Since the trend followers induce positive autocorrelations and the value investors negative autocorrelations, there is predict-

13. The smoothed volume is computed as $\bar{V}_t = \beta \bar{V}_{t-1} + (1-\beta) V_t$, where $V_t$ is the volume and $\beta = 0.9$.

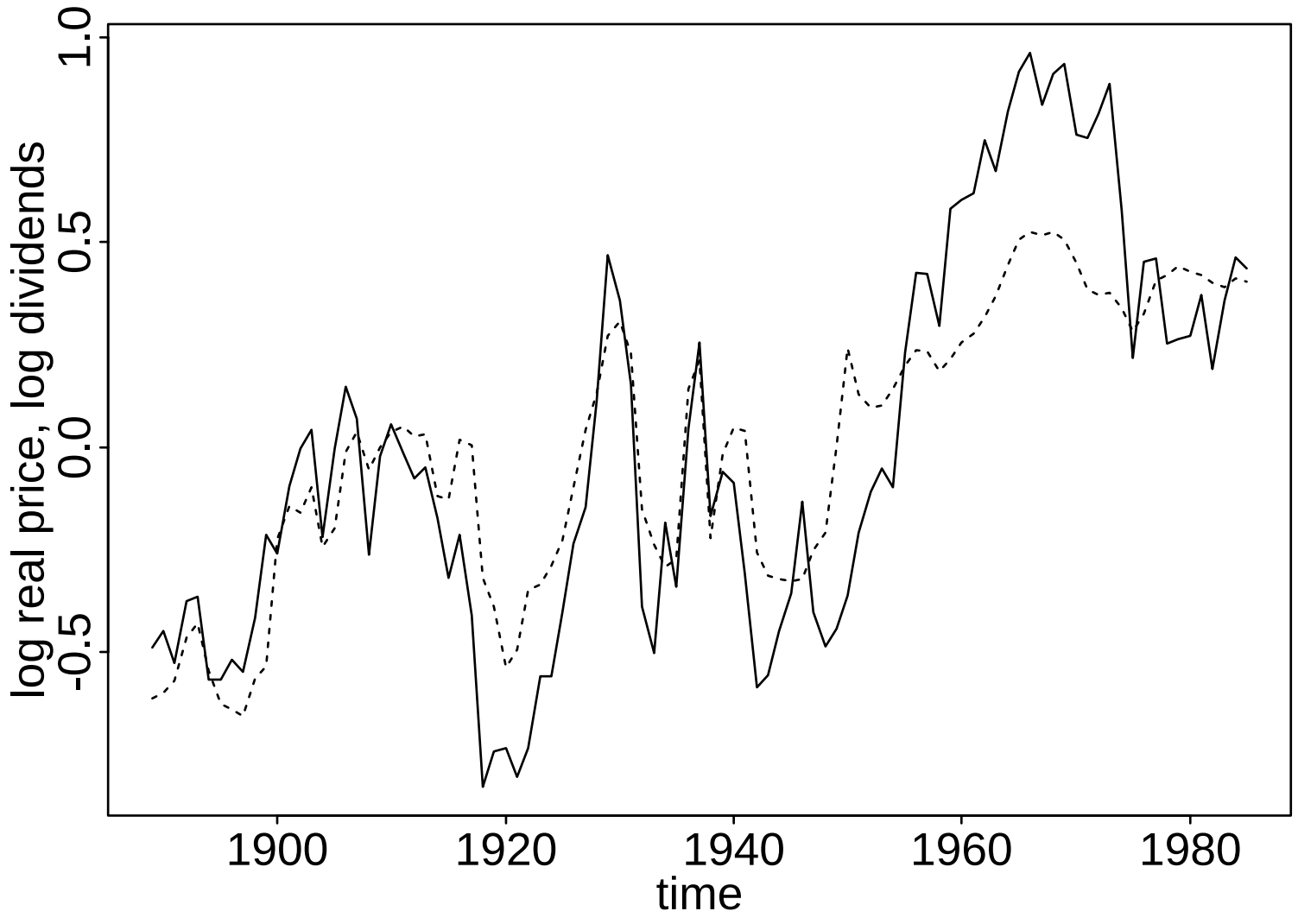

**FIGURE 9. Inflation-adjusted annual prices (solid) and dividends for the S&P index of American stock prices.**

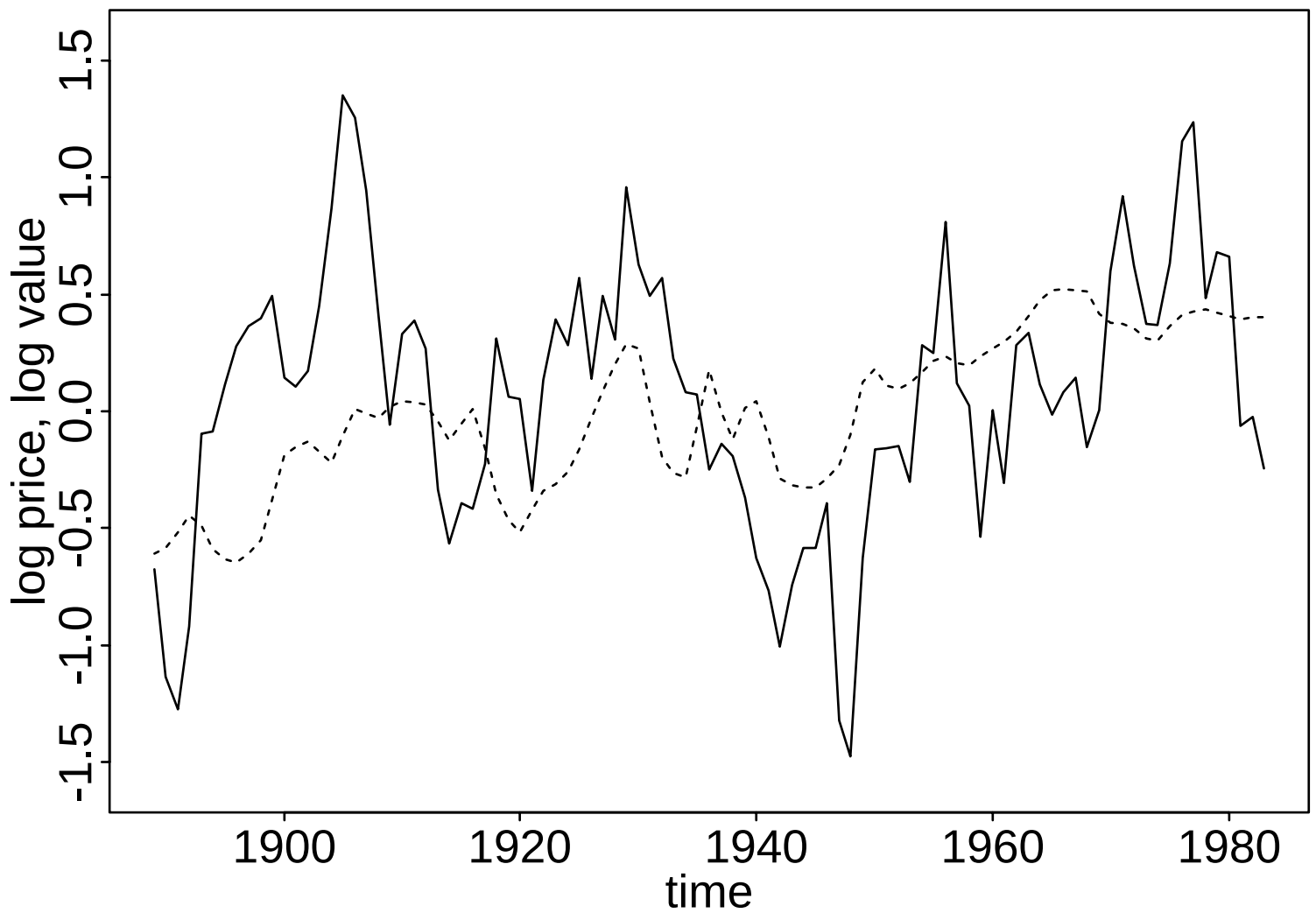

**FIGURE 10. A simulation with value investors and trend followers. The linearly interpolated dividend series from Figure 9 provides the reference value process. Prices are averaged to simulate reduction to annual data. There was some adjustment of parameters, as described in the text, but no attempt was made to match initial conditions. The oscillation of prices around values is qualitatively similar to Figure 9.**

able nonlinear structure for a trader who understands the underlying dynamics well enough to predict which group will become active. Without knowledge of the underlying generating process, however, it is difficult to find such a forecasting model directly from the timeseries.

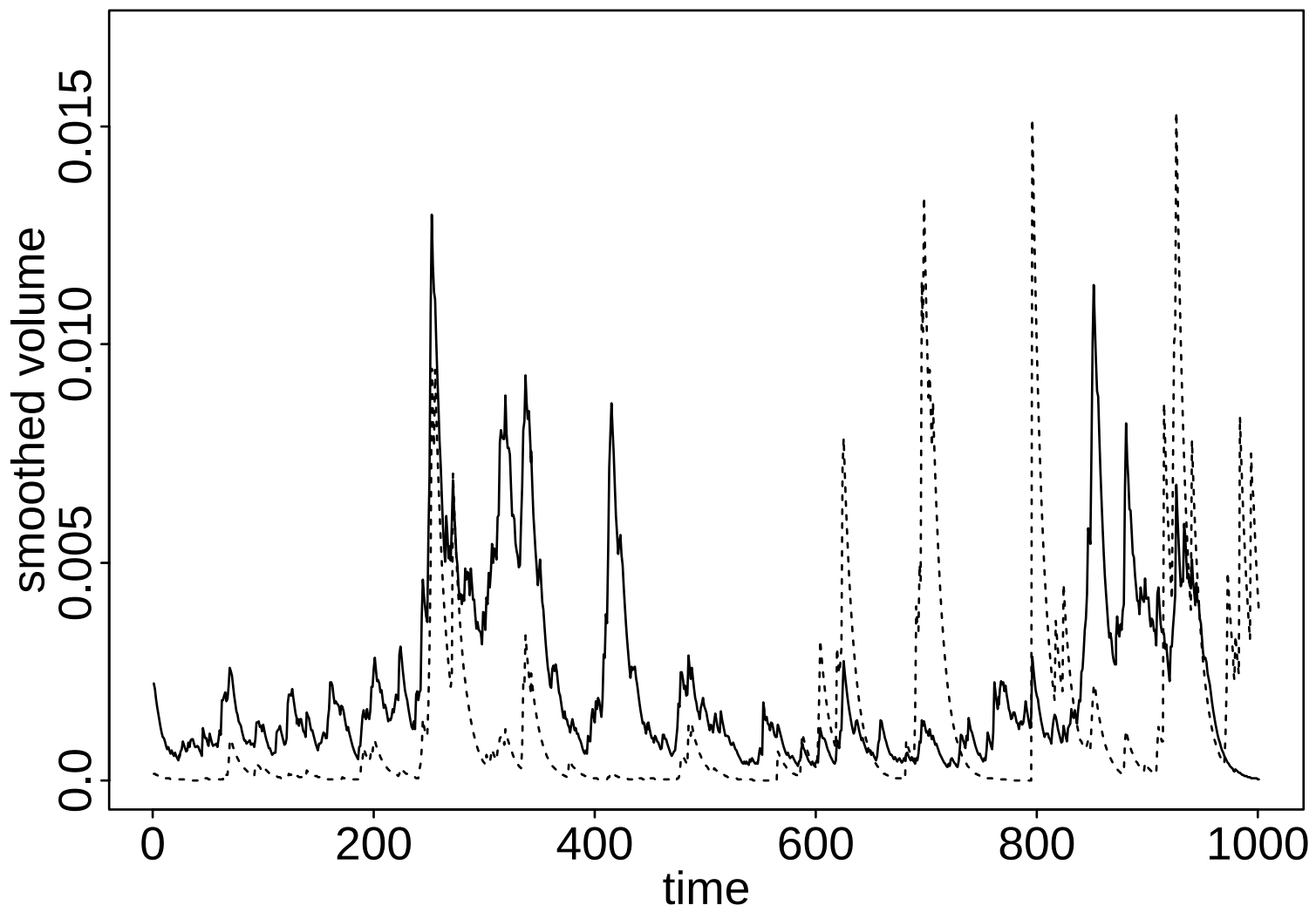

**FIGURE 11. Smoothed trading volume of value investors (solid line) and trend followers (dashed line). The two groups become active at different times; when the value investors dominate the log-returns have a negative autocorrelation, and when the trend followers dominate there is a positive autocorrelation. Even though there is no linear temporal structure, there is strong nonlinear structure. Parameters are as described in Table 1; this is only a short portion of the total simulation.**

Statistical analyses display many of the characteristic properties of real financial timeseries, as illustrated in Figure 12. The log-returns are more long-tailed than those of a normal distribution, i.e. there is a higher density of values at the extremes and in the center with a deficit in between. This also evident in the size of the fourth moment. The excess kurtosis $k = \langle (r_t - \bar{r}_t)^4 \rangle / \sigma_r^4 - 3$ is roughly $k \approx 9$, in contrast to $k = 0$ for a normal distribution. The histogram of volumes is peaked near zero with a heavy positive skew. The volume and volatility both have strong positive autocorrelations. The intensity of the long-tails and correlations vary as the parameters are changed or strategies are altered. However, the basic properties of long tails and autocorrelated volume and volatility are robust as long as trend followers are included.

Clustered volatility has now been seen in many different agent-based models[14]. It seems there are many ways to do produce this behavior. The mechanism in this case is due to positive feedback: Large price fluctuations cause large trading volume, which causes large price fluctuations, and so on, generating volatility bursts. Even without any autocorrelations in prices themselves, the nonlinearly driven variations in the trading activity of value and trend strategies can cause autocorrelations in volatility. Several authors, including Lux et al. (1997, 1998, 1999) and Brock and Hommes (1997, 1998, 1999) have suggested that fluctuating volatility is driven by changes in the population of trend followers.

For real markets, this may be problematic: Real agents may not change strategies this fast. The feedback hypothesis offered here does not require agents to change strategies. However, it is not clear whether the resulting volatility correlations are strong enough to match those observed in real data. More work is needed to resolve this question.

The few results presented here fail to do justice to the richness of the trend follower/value investor dynamics. We have observed many interesting effects. For example, the presence of trend followers increases the frequency of oscillations in mispricing. The mechanism seems to be more or less as follows: If a substantial mispricing develops by chance, value investors become active. Their trading shrinks the mispricing, with a corresponding change in price. This causes trend followers to become active; first the short term trend followers enter, and then successively longer term trend followers enter, sustaining the trend and causing the mispricing to cross through zero. This continues until the mispricing becomes large, but with the opposite sign, and the process repeats itself. As a result the oscillations in the mispricing are faster than they would be without the trend followers. This mechanism is a less regular version of that postulated by Chiarella (1992).

## 4. Concluding remarks

These results illustrate how commonly used trading strategies can be viewed as signal processing elements. Trend following strategies act as signal filters, amplifying high frequency noise and inducing short term positive autocorrelations. Value investing strategies act as signal transducers, incorporating information about value into prices, and inducing negative short term autocorrelations. The fact that prices have very small autocorrelations suggests that value investors alone cannot be the only group present -- there must be other groups present, such as trend followers, to cancel their negative autocorrelations.

Nonlinear value investing strategies can amplify noise in a heterogeneous setting where there are diverse views concerning value. Trend following strategies strongly amplify high frequency noise, so that when the two groups are combined the result is excess volatility. When value investing and trend following strategies are combined, by adjusting their relative populations, the short term autocorrelations can be made to cancel, so that in a long time average there is very little linear structure. However, because each style of trading is activated differently, there may be bursts of trading by either group, even without agents defecting from one group to the other. The feedback effects studied here give rise to clustered volatility; unlike explanations that rely on oscillations in the populations of different groups of traders, this explanation is plausible even on fairly rapid times-cales. Whether such feedback effects are strong enough to explain the clustered volatility observed in real markets, however, remains an open question.

---

14. Some examples include Brock and LeBaron (1996), Levy et al. (1996), Takayasu et al. (1997), Arthur et al. (1997), LeBaron et al. (1999), Caldarelli et al. (1997), Brock and Hommes (1997, 1998, 1999), Lux (1997,1998), Lux and Marchesi (1999), Youssefmir et al.(1998), Bouchaud and Cont (1998), Gaunersdorfer and Hommes (1999) and Iori (1999). Fat tails with realistic tail exponents have been observed by Lux and Marchesi (1999) in simulations of value investors and trend-followers based on the log-linear price formation rule; Stauffer and Sornette (1999) have predicted realistic exponents using equation 8 with randomly varying liquidity.

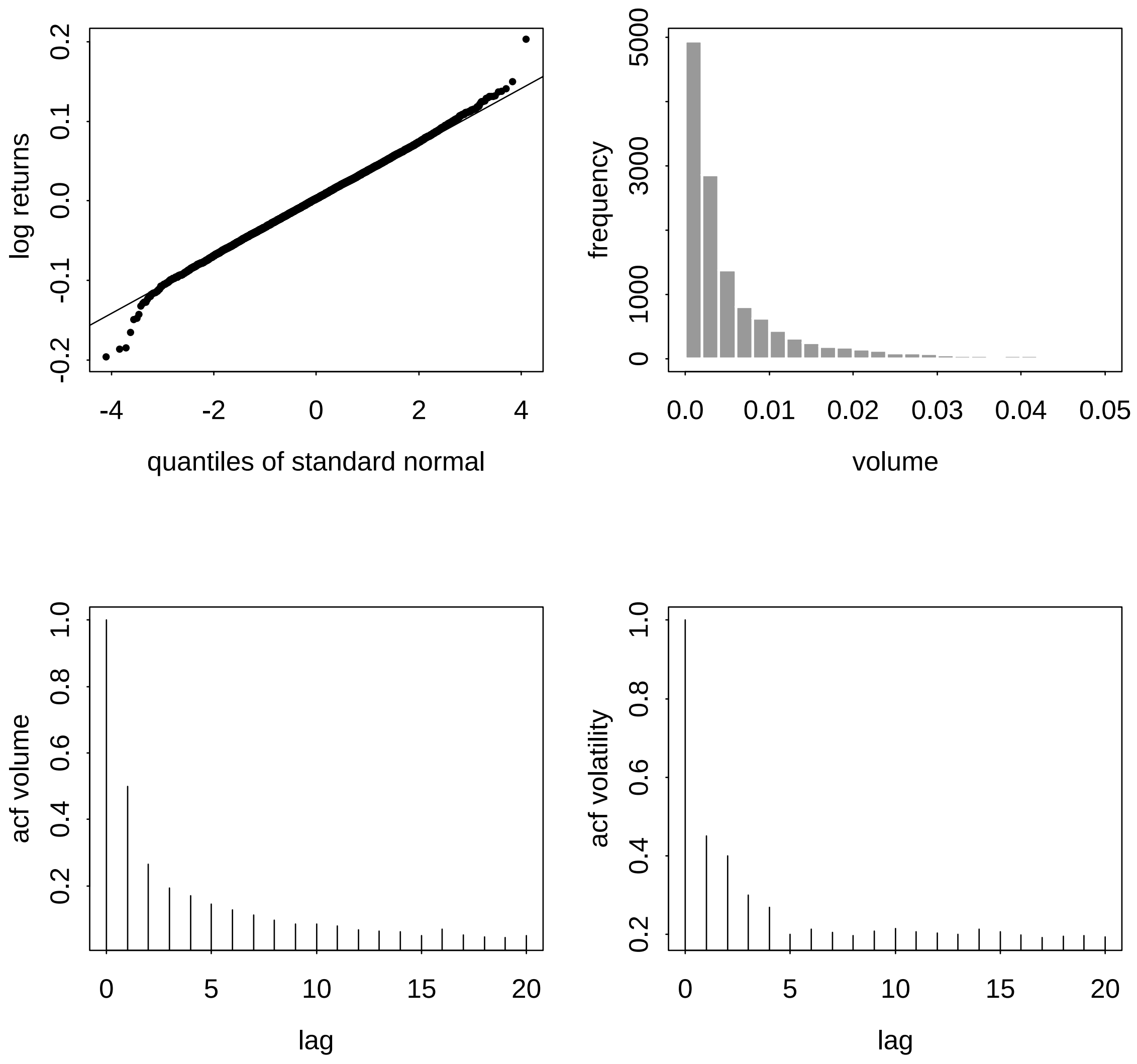

**FIGURE 12. An illustration that an ecology of threshold based value investors and trend followers shows statistical properties that are typical of real financial time series. The upper left panel is a "q-q" plot, giving the ratio of the quantiles of the cumulative probability distribution for the log-returns to those of a normal distribution. If the distribution were normal this would be a straight line, but since it is "fat tailed" the slope is flatter in the middle and steeper at the extremes. The upper right panel shows a histogram of the volume. It is heavily positively skewed. The lower left panel shows the autocorrelation of the volume, and the lower right panel shows the autocorrelation of the volatility. These vary based on parameters, but fat tails and temporal autocorrelation of volume and volatility are typical.**

A key element missing from the price formation mechanism studied here is risk aversion by the market maker. This has several profound effects on price dynamics. First, it serves to reduce deviations from market clearing, and makes prices track values more closely. However, it also induces additional temporal structure in prices. Under the price formation mechanism of equation 8, for example, the market maker's price adjustments are positively correlated. This can be exploited by trend followers, and provides one possible explanation for the persistence of trend followers. In a future paper we will present some results that include market maker risk aversion, and which also study the profitability and reinvestment dynamics of different groups of agents.

## Acknowledgments

We would like to thank Paul Melby for contributing Figure 3, and John Geanakoplos for helpful discussions. We also thank the McKinsey Corporation for their generous support.